\documentclass[aps,prb,twocolumn]{revtex4}
\usepackage{amsmath} 
\usepackage{amsfonts} 
\usepackage{graphicx} 

\begin{document}
\title{Slow-dissipation limit of the harmonic oscillator with general power-law damping}
\author{Jarrett L. Lancaster}
\email[Electronic mail: ]{jlancas2@highpoint.edu}
\affiliation{Department of Physics, High Point University, One University Parkway, High Point, NC 27268} 
\date{\today}

\begin{abstract}
An approximate solution is presented for simple harmonic motion in the presence of damping by a force which is a general power-law function of the velocity. The approximation is shown to be quite robust, allowing for a simple way to investigate amplitude decay in the presence of general types of weak, nonlinear damping. \end{abstract}
\maketitle
A standard rite of passage for the undergraduate mechanics student is exploring the solution to the harmonic oscillator in the presence of linear damping,
\begin{equation}
m\ddot{x} + kx - f_{\alpha}(\dot{x}) = 0,\label{eq:sho}
\end{equation}
for which $f_{\alpha}(\dot{x})\rightarrow f_{1}(\dot{x}) = -b\dot{x}$. Given initial conditions $x(0) = x_{0}$ and $\dot{x}(0) = v_{0}$, the underdamped $\left(\frac{b}{2m}<\omega_{0}\right)$ solution takes the form\cite{McCormack}
\begin{equation}
x(t) = e^{-\gamma t}\left[x_{0}\cos\left[\omega t\right] + \left(\frac{v_{0} + \gamma x_{0}}{\omega}\right)\sin\left[\omega t\right] \right],\label{eq:shoexact}
\end{equation}
where $\gamma \equiv \frac{b}{2m}$, $\omega \equiv \sqrt{\omega_{0}^{2}-\gamma^{2}}$ and $\omega_{0}^{2} \equiv \frac{k}{m}$. In the {\it weak-damping} limit $\gamma \ll \omega_{0}$ when $v_{0} = 0$, the solution simplifies substantially to
\begin{equation}
x(t) \simeq  x_{0}e^{-\gamma t}\cos\left[\omega_{0}t\right]\;\;\;\;\;\;(\gamma \ll \omega_{0}).\label{eq:approx1}
\end{equation}
Here Eq.~(\ref{eq:approx1}) has the form $x(t) \simeq A(t)\cos(\omega_{0}t)$ with $A(t) = x_{0}e^{-\gamma t}$, corresponding to the undamped ($\gamma = 0$) solution with a time-dependent amplitude. In the remainder of this note, this approximate form is generalized for the case of arbitrary power-law damping functions, $f_{\alpha}(\dot{x}) = -b\dot{x}|\dot{x}|^{\alpha-1}$ and the condition which generalizes the linear, ``weak-damping'' limit $\gamma \ll \omega_{0}$ is discussed.

Students and instructors might wonder why linear damping is the only case investigated in detail in most texts on advanced, undergraduate-level, classical mechanics.\cite{Taylor,Marion,Fowles} The exclusive focus of most authors on linear, viscous damping is likely due to the fact that Eq.~(\ref{eq:sho}) is a linear equation--and therefore solvable in terms of elementary functions--only for the case $\alpha = 1$. It is worth noting that other values of $\alpha$ also lead to physically plausible dissipative forces, since $\alpha = 0$ represents a constant force of ``dry,'' sliding friction\cite{Barratt,Beichner,DeAmbrosis,Hauko} while $\alpha = 2$ corresponds to turbulent air resistance.\cite{White,Smith} Additionally, a cubic drag term, which has attracted significant attention as a nontrivial mathematical problem,\cite{Kahn} has appeared as an important feature in studies of energy-harvesting devices and cochlear implants.\cite{Elliott} Large values of $\alpha >2$ have also been explored as drag terms for the nonlinear pendulum.\cite{Quiroga}

As no fundamental law forbids non-integer values of $\alpha$, it is perhaps unsurprising that non-integer values of $\alpha$ also find applications. The effective drag force on objects moving through fluids\cite{White} sometimes obeys such a nonintegral power law experimentally.\cite{Terenzi} Fractional power-law functions also appear as relaxation functions in certain models of viscoelasticity\cite{Pritchard} relevant to the study of biological tissues. Such systems are substantially more complicated than harmonic oscillators. Still, aside from the purely mathematical curiosity of nonlinear damping, intuition gleaned from investigating an oscillator with nonlinear damping could be useful to students who wish to explore viscoelasticity and its role in biomechanics. The aim of this note is to summarize a unified, analytic treatment of the asymptotic behavior predicted by Eq.~(\ref{eq:sho}) for all possible values of $\alpha$. Additionally, we consider briefly the situation in which more than one damping term is present. The general approach employed throughout this note serves as a useful example of how one may extract meaningful predictions from analytically intractable systems by using asymptotic methods. Few interesting problems in physics are exactly solvable, and students should benefit from early exposure to such methods in a familiar setting such as the harmonic oscillator.

For an oscillator obeying Eq.~(\ref{eq:sho}) with arbitrary $\alpha$, one may write the exact solution as $x(t) = A(t)\cos[\omega_{0}t+\varphi(t)]$. We take $\varphi(0) = 0$, corresponding to $\dot{x}(0) = 0$ without loss of generality. For {\it weak} damping, it is physically plausible to suppose $\varphi(t) = \varphi(0)$, but this point will be revisited below. Supposing the phase remains equal to its initial value, the energy contained in the oscillator at some time $t$ is
\begin{eqnarray}
E(t) & = & \frac{1}{2}m\dot{x}^{2}(t) + \frac{1}{2}kx^{2}(t),\label{eq:E1}\\
& = & \frac{1}{2}kA^{2}(t)\left[1 + \left(\frac{\dot{A}(t)}{\omega_{0}A(t)}\right)^{2}\cos^{2}(\omega_{0}t)\right.\nonumber\\
& &  - \left.\left(\frac{\dot{A}(t)}{\omega_{0}A(t)}\right)\sin(2\omega_{0}t)\right],
\end{eqnarray}
where the second line follows from evaluating $\dot{x} = \dot{A}(t)\cos(\omega_{0}t) - A(t)\omega_{0}\sin(\omega_{0}t)$. We consider the {\it slow dissipation} limit for which $\dot{A} \ll \omega_{0}A$, corresponding to a regime in which the timescale over which the amplitude changes appreciably is much longer than the period of oscillation. Neglecting contributions involving $\dot{A} \ll \omega_{0}A$ gives
\begin{equation}
E \simeq \frac{1}{2}kA^{2}(t).\label{eq:E2}
\end{equation}
Differentiating both sides of Eq.~(\ref{eq:E1}) with respect to time gives 
\begin{equation}
\dot{E} = \dot{x}\left[m\ddot{x} + kx\right]= f_{\alpha}(\dot{x})\dot{x},
\end{equation} 
where the last line follows from using Eq.~(\ref{eq:sho}). Taking the time derivative of Eq.~(\ref{eq:E2}) gives $\dot{E} = kA\dot{A}$ and equating the two expressions for $\dot{E}$ yields the relationship
\begin{equation}
kA\dot{A} = f_{\alpha}(\dot{x})\dot{x}.\label{eq:ode1}
\end{equation}
Equation~(\ref{eq:ode1}) becomes quite useful in the limit of slow dissipation ($\dot{A} \ll \omega_{0}A$), where 
\begin{equation}
\dot{x}(t) \approx -A\omega_{0}\sin[\omega_{0}t].
\end{equation}
Using Eq.~(\ref{eq:ode1}) and taking $f_{\alpha}(\dot{x}) = -b\dot{x}|\dot{x}|^{\alpha-1}$, one may separate variables and integrate. The basic approach described here was employed previously in Ref.~\onlinecite{Wang} where the exponents $\alpha =0,1,2$ were treated as separate cases. The aim of the present note is to demonstrate that these cases can be obtained as limits of a more general solution for arbitrary $\alpha$ while also exploring the domain of validity for the approximations used. Indeed, this general method of solution is a particularly intuitive application of {\it multiple scale analysis},\cite{Kahn,Bender} which is essentially an expansion of the solution at long times in powers of the (dimensionless) damping coefficient $\frac{b\omega_{0}^{\alpha}}{kx_{0}^{1-\alpha}}$. It can be shown that the phase remains equal to its initial value (here zero) to lowest order in $b$. Indeed, the case $\alpha = 3$ was considered as an illustrative example of this method in Ref.~\onlinecite{Bender}. Taking $A(0) = x_{0}$, the result is
\begin{eqnarray}
\frac{A(t)^{1-\alpha}-x_{0}^{1-\alpha}}{1-\alpha} = -\frac{b\omega_{0}^{1+\alpha}}{k}\int_{0}^{t}|\sin^{1+\alpha}(\omega_{0}t')|dt' \nonumber\\
\;\;\;\;\;\;\;\;\;\;\;\;\;\;\;(\alpha \neq 1).\label{eq:alphan1}
\end{eqnarray}
The case $\alpha = 1$, corresponding to linear damping, follows from writing $A^{1-\alpha} = e^{(1-\alpha)\ln A}$ and expanding the exponential for small $\epsilon = 1-\alpha$ so that the limit $\epsilon\rightarrow 0$ may be taken. The remaining work concerns the integral appearing on the right-hand sides of Eq.~(\ref{eq:alphan1}). Due to the periodic, non-negative nature of the integrand, the result should be approximately proportional to time $t$. Therefore we may write
\begin{equation}
\int_{0}^{t}|\sin^{\alpha+1}(\omega_{0}t')|dt' \simeq \left\langle \left|\sin^{1+\alpha}\left(\omega't\right)\right|\right\rangle t,\label{eq:tav1}
\end{equation}
where $\left\langle f(t) \right\rangle$ denotes the average value of a function $f$. For a {\it periodic} function $f(t)$, the average only needs to be computed over a single period. In the case of $|\sin^{\alpha+1}(\omega_{0}t)|$, the period is $\frac{\pi}{\omega_{0}}$, and this average can be evaluated in terms of the Euler beta function\cite{Gradshteyn}
\begin{equation}
\int_{0}^{t}|\sin^{\alpha+1}(\omega t')|dt'  \simeq 2^{\alpha+1}B\left(\frac{\alpha}{2}+1,\frac{\alpha}{2}+1\right) t/\pi.\label{eq:tav2}
\end{equation}
where $B(x,y) \equiv   \Gamma(x)\Gamma(y)/\Gamma(x+y)$. For $\alpha \neq 1$, the result is
\begin{eqnarray}
A(t) & = & x_{0}\left[1+(\alpha-1)\frac{b(2\omega_{0})^{\alpha+1}t}{\pi k x_{0}^{1-\alpha}}B\left[\frac{\alpha}{2}+1,\frac{\alpha}{2}+1\right]\right]^{-\frac{1}{\alpha-1}} \nonumber\\
&&\;\;\;\;\;\;\;\;\;\;\;\;\;\;\;\;\;\;\;\;\;\;\;\;\;\;\;\;\;\;\;\;\;\;\;\;\;\;\;\;\;\;\;\;\;\;(\alpha \neq 1).\label{eq:At}
\end{eqnarray}

The particular expressions which follow for $\alpha = 0,1,\frac{3}{2},2,3$ are collected in Table~\ref{tab:tab1}. For $\alpha >1$, the dependence of the amplitude on the initial displacement $x_{0}$ disappears in the long-time limit, as is known to be the case for quadratic damping.\cite{Zeldovich} As noted previously, the cases for $\alpha = 0,1,2$ have been obtained in Ref.~\onlinecite{Wang} by working out the integral in Eq.~(\ref{eq:tav2}) separately for each case. Note that Eq.~(\ref{eq:At}) contains all of this information as specific cases, and the mathematically brave student can capture all three from a single integration of Eq.~(\ref{eq:alphan1}). With minimal additional effort, one may also obtain $A(t)$ for any real value of damping exponent with $\alpha = \frac{3}{2},3$ depicted as specific examples. It has been claimed\cite{Quiroga} that the cases $\alpha >2$ are not physical, as the amplitude decays too slowly to agree with experimental systems. Indeed, the predicted decay becomes slower with increasing $\alpha$, since Eq.~(\ref{eq:At}) shows $A(t) \sim t^{-\frac{1}{\alpha -1}}$ as $t\rightarrow \infty$. The validity of this claim will be revisited below when multiple damping terms are included.

\begin{table}[h]
\caption{Approximate time-dependence of oscillation amplitude in low-dissipation limit computed form Eq.~(\ref{eq:At}) for $\alpha=0,1,\frac{3}{2},2,3$}
\centering
\begin{tabular}{l c c} 
\hline\hline 
$\alpha$  & \;\;\;\;\;\;\;\;\;\;\;\;\;\;\;\;\;\;\;\;\;\;\;\;\;\;\;\;\;\;\;\;\;\;\;\;\;\;\;\;\;\; & $A(t)$\\ [0.5ex]
\hline\\ [.5ex] 
$0$ & & $\displaystyle x_{0} - \frac{2bt}{\pi m\omega_{0}}$\\ [2.5ex]
$1$ & & $\displaystyle x_{0}e^{-\frac{bt}{2m}}$\\ [2.5ex]
$\frac{3}{2}$ & & $\displaystyle x_{0}\left[1+ \frac{3bt\sqrt{2x_{0}\omega_{0}} \Gamma^{2}\left(\frac{3}{4}\right)}{5\pi^{\frac{3}{2}} m}\right]^{-2}$\\ [4.5ex]
$2$ & & $\displaystyle \frac{x_{0}}{1+ \displaystyle\frac{4b\omega_{0}x_{0}t}{3\pi m}}$\\[5.5ex]
$3$ && $\displaystyle \frac{x_{0}}{\sqrt{1 + \frac{3bkx_{0}^{2}t}{4m^{2}}}}$\\[5.5ex]
\hline\hline 
\end{tabular}
\label{tab:tab1}
\end{table}
Using Eq.~(\ref{eq:At}), we can explore the domain of validity for the slow-dissipation limit with arbitrary damping. The (approximate) ansatz $x(t) \simeq A(t)\cos(\omega_{0}t)$ was inspired by the {\it underdamped} solution corresponding to $\alpha = 1$ and could be viewed as the {\it weak-damping} case for the linear oscillator. Here we shall see that the restriction of weak damping is generally {\it not} equivalent to that of slow dissipation. For $\alpha \neq 1$, the approximation will work best at either long or short times. To satisfy $\dot{A}\ll\omega_{0}A$ for general $\alpha\neq1$, one must have
\begin{equation}
\frac{|\dot{A}|}{A} = \kappa\omega_{0}\left|1+(\alpha-1)\kappa \omega_{0} t\right|^{-1} \ll \omega_{0},\label{eq:wd}
\end{equation}
where $\kappa \equiv 2b(2\omega_{0})^{\alpha}B\left[\frac{\alpha}{2}+1,\frac{\alpha}{2}+1\right]/(\pi kx_{0}^{1-\alpha})$ is a dimensionless parameter. The case $\alpha = 1$ reduces directly to $\frac{b}{2m}\ll \omega_{0}$, indicating the equivalence of the slow-dissipation limit to the weak-damping limit. For $\alpha > 1$, Eq.~(\ref{eq:wd}) is satisfied in the long-time limit where $(\alpha-1)\kappa\omega_{0}t\gg1$ so that the condition reduces simply to 
\begin{equation}
\frac{|\dot{A}|}{A} \xrightarrow{\omega_{0}t\gg1} \frac{1}{(\alpha-1)t} \ll \omega_{0}.
\end{equation}
That is, Eq.~(\ref{eq:At}) becomes valid at times $t \gg \frac{1}{(\alpha-1)\omega_{0}}$ {\it regardless} of the value of $b$. An interesting corollary to this statement is that critically damped or overdamped behavior does not occur for any damping strength when $\alpha >1$. The absence of critically damped or overdamped dynamics has been demonstrated previously\cite{Smith} for the particular case of $\alpha =2$. The conjecture that only underdamped behavior occurs for $\alpha>1$ is somewhat speculative, as the asymptotic results are formally correct only to leading order in a hypothetical expansion in powers of the dimensionless parameter $\kappa\propto b$. Agreement between the asymptotic results and a numerical integration (see below) turns out to be quite good, even for large values of $b$ with only the phase of the asymptotic solution disagreeing substantially from the numerical solution. It is interesting to note that the fractional Maxwell model for viscoelastic materials also shows an absence of overdamped and critically damped behavior away from the linear regime.\cite{Pritchard} However, models of viscoelasticity are considerably more complex than those of damped harmonic oscillators, so there is little room for a true analogy. 
\begin{equation}
\frac{b}{kx_{0}} = \frac{F_{\mbox{\tiny friction}}}{F^{(\mbox{\tiny max})}_{\mbox{\tiny spring}}} \ll 1.
\end{equation}
\begin{figure}
\includegraphics[totalheight=6.5cm,]{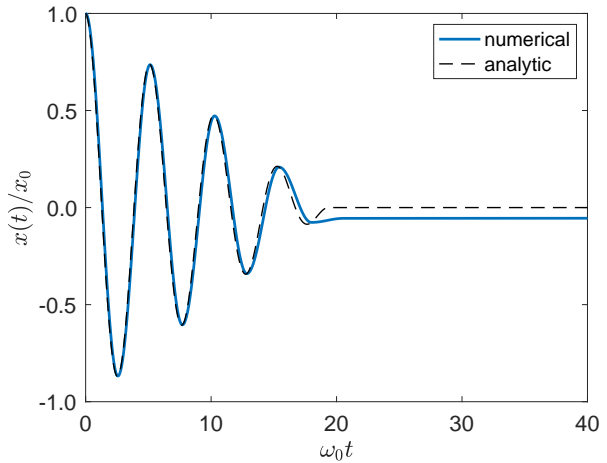}
\caption{Comparison of numerical solution of Eq.~(\ref{eq:sho}) with $\alpha = 0$ to analytic approximation $x_{a}(t) = A(t)\cos(\omega_{0}t)$ with $A(t)$ given in Table~\ref{tab:tab1}. Here $\kappa \approx 0.042$, so that according to Eq.~(\ref{eq:wd2}) the approximation should work well at short times.  Sliding friction allows for the mass to stop away from equilibrium, which is not captured by the approximate solution.}
\label{fig:plot0}
\end{figure}
\begin{figure}
\includegraphics[totalheight=6.5cm,]{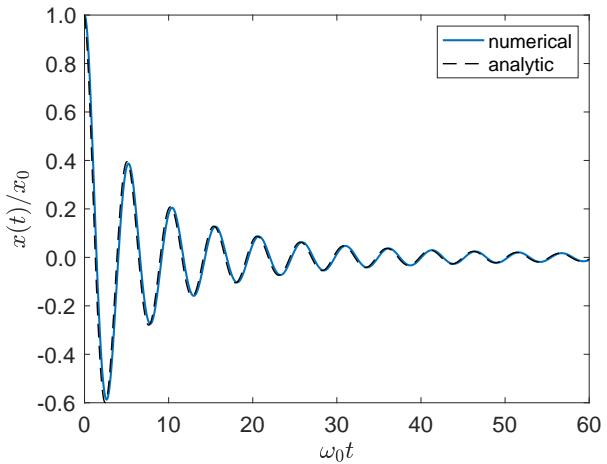}
\caption{Comparison of numerical solution of Eq.~(\ref{eq:sho}) with $\alpha = \frac{3}{2}$ to analytic approximation $x_{a}(t) = A(t)\cos(\omega_{0}t)$ with $A(t)$ given in Table~\ref{tab:tab1}. Here $\kappa \approx 0.19$. The case shown is fairly generic for $\alpha >1$ in which the approximation is quite robust, even for modest $\kappa$.}
\label{fig:plot15}
\end{figure}
Interestingly, the analytically solvable case $\alpha =1$ represents a sort of crossover case between where slow-dissipation holds at long-times and where it holds at short times. For $\alpha <1$, the constant $\kappa$ scales with the physical constants as $\kappa \sim \frac{b\omega_{0}^{\alpha}}{k x_{0}^{1-\alpha}}$, and the slow-dissipation condition yields
\begin{equation}
\frac{|\dot{A}|}{A} \sim \frac{b\omega_{0}^{\alpha+1}}{kx_{0}^{1-\alpha}}\left|1 -(1-\alpha)\frac{b\omega_{0}^{\alpha+1}}{kx_{0}^{1-\alpha}}t\right|^{-1} \ll\omega_{0}.\label{eq:wd2}
\end{equation}
Equation~(\ref{eq:wd2}) is less transparent than the case for $\alpha>1$, but at short times it is clear that slow dissipation only occurs when certain conditions are met on damping strength $b$, initial amplitude $x_{0}$, angular frequency $\omega_{0}$ and spring stiffness $k$. ``Weak damping'' corresponds to $b \ll \frac{kx_{0}^{1-\alpha}}{\omega_{0}^{\alpha}}$, or $\kappa \ll1$. Indeed, for the case of sliding friction ($\alpha =0$), $b$ represents the constant sliding friction force, and the weak damping simply means the frictional force is small compared the the scale of the spring's restoring force,

However, even if this weak-damping condition is satisfied, Eq.~(\ref{eq:wd2}) cannot be satisfied for {\it long} times $\omega_{0}t\gg1$ since the denominator shrinks with increasing time. Thus for $\alpha <1$, the asymptotic results for $A(t)$ are valid for the weak-damping regime {\it at short times}. Figure~\ref{fig:plot0} depicts the gradual breakdown for the case $\alpha = 0$. For certain parameter choices, it is possible for the sliding friction to overwhelm the restoring force of the spring when the amplitude has decayed to a sufficiently small value. Generally, one has $\mu_{s}>\mu_{k}$ for static ($\mu_{s}$) and kinetic ($\mu_{k}$) coefficients of friction, so this behavior should be treated more carefully for more physically correct results at long times.\cite{note4}  Effectively, the model considered treats $\mu_{s} = \mu_{k} = \frac{b}{mg}$, and our main interest is in demonstrating the validity of the asymptotic result at short times. It should be noted that examples in which the mass comes to rest away from equilibrium have been published for exact solutions\cite{Beichner} as well as experimental realizations.\cite{DeAmbrosis,Hauko} For $\alpha >1$ and modest values of $\kappa$, agreement between numerical and approximate solutions at long times $\omega_{0}t \gg 1$ is generally quite good, as shown in Fig.~\ref{fig:plot15}. A \textsc{Python} implementation of the numerical approach used to generate these figures is included as supplemental information.\cite{code}

Lastly, we consider briefly the possibility of two distinct types of damping. That is, Eq.~(\ref{eq:ode1}) is modified as
\begin{eqnarray}
kA\dot{A} & = & -b \dot{x}^{2}|\dot{x}|^{\alpha-1} - b'\dot{x}^{2}|\dot{x}|^{\alpha'-1},\\
& \simeq & -b A^{\alpha + 1}\omega_{0}^{\alpha + 1}\left|\sin^{\alpha + 1}(\omega_{0}t)\right| \nonumber\\
& & - b' A^{\alpha' + 1}\omega_{0}^{\alpha ' + 1}\left|\sin^{\alpha' + 1}(\omega_{0}t)\right|,
\end{eqnarray}
for two power-law functions with exponents $\alpha$, $\alpha'$ and respective amplitudes $b$, $b'$. Without loss of generality, we take $\alpha' >\alpha$. In the second line, the approximation $\dot{x}(t) \simeq \omega_{0}A\sin(\omega_{0}t)$ has been applied. For $\alpha \neq \alpha '$, the approximate differential equation governing $A(t)$ is no longer separable. However, the heart of the approximation used in the simpler case was to replace oscillatory terms of the form $|\sin^{\alpha + 1}(\omega_{0}t)|$ by their {\it average} values over a single oscillation cycle. Making this approximation and using Eqs.~(\ref{eq:tav1})--(\ref{eq:tav2}) yields a separable, approximate equation for $A(t)$,
\begin{equation}
\int_{x_{0}}^{A(t)}\frac{dA}{cA^{\alpha} + c'A^{\alpha'}} \simeq -\frac{t}{\pi k},\label{eq:twodrag}
\end{equation}
where $c = b (2\omega_{0})^{\alpha + 1}B\left(\frac{\alpha}{2} + 1,\frac{\alpha}{2} + 1\right)$ and $c' = b' (2\omega_{0})^{\alpha' + 1}B\left(\frac{\alpha'}{2} + 1,\frac{\alpha'}{2} + 1\right)$. The general case does not lend itself to transparent results,\cite{note3} so let us take the case of $\alpha = 1$ and $\alpha' = 2$ corresponding to a combination of viscous drag and air resistance. In this case the integral in Eq.~(\ref{eq:twodrag}) is straightforwardly evaluated, giving
\begin{equation}
A(t) \simeq \frac{c\lambda e^{-\frac{ct}{\pi k}}}{1-\lambda c' e^{-\frac{ct}{\pi k}}},\label{eq:atwodrag}
\end{equation}
where $\lambda \equiv \frac{x_{0}}{c+c'x_{0}}$. For $\alpha =1$ and $\alpha'=2$, one has $c = \frac{\pi b k}{2m}$ and $c' = \frac{4b'\omega_{0}k}{3m}$. In the long time limit the exponential in the denominator of Eq.~(\ref{eq:atwodrag}) may be neglected, so that the asymptotic form of the amplitude may be written
\begin{equation}
A(t)  \xrightarrow{\omega_{0}t \gg 1} \frac{x_{0}}{1 + \frac{8b'\omega_{0}}{3b \pi}x_{0}}e^{-\frac{b t}{2m}}.\label{eq:atwodrag2}
\end{equation}
\begin{figure}
\includegraphics[totalheight=6.5cm,]{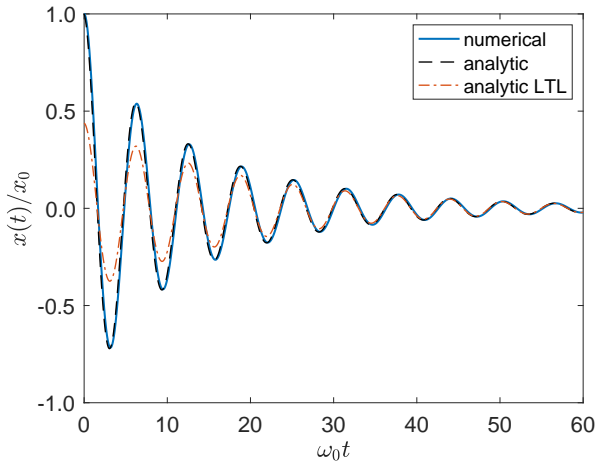}
\caption{Comparison of numerical solution of Eq.~(\ref{eq:sho}) with combined total damping force $-b\dot{x}|\dot{x}|^{\alpha-1}-b'\dot{x}|\dot{x}|^{\alpha'-1}$ for $\alpha=1$, $\alpha' = 2$ and $b=\omega_{0}x_{0}b'$. The analytic expression $A(t)\cos[\omega_{0}t]$ with $A(t)$ given by Eq.~(\ref{eq:atwodrag}) provides an excellent approximation in the slow-dissipation regime. At long times, the simpler long-time limit (LTL) in Eq.~(\ref{eq:atwodrag2}) becomes an accurate representation.}
\label{fig:plot3}
\end{figure}
A few general features can be gleaned from Eq.~(\ref{eq:atwodrag2}). First, the long-time decay envelope controlled by the exponential factor $e^{-\frac{b t}{2m}}$ which also describes the case of only linear damping. Rather than some crossover/intermediate behavior, we find the long-time decay is dominated by the term with the smallest value of $\alpha$. So far we have made no restrictions whatsoever on the relative sizes of $b$ and $\omega_{0}x_{0}b'$.\cite{note5} In the limit of negligible quadratic damping, $\omega_{0}x_{0}b' \ll b$, the second term in the denominator may be neglected so one recovers the standard result for linear damping, $A(t)\rightarrow x_{0}e^{-\frac{b t}{2m}}$. More interesting is the case that $\omega_{0}x_{0}b' \gg b$ so that the quadratic drag is ``stronger.'' In this limit $A(t)\rightarrow \frac{3b \pi}{8b' \omega_{0}}e^{-\frac{b t}{2m}}$. While the decay at long times is still dominated by the exponential envelope characteristic of linear drag, the dependence of the instantaneous amplitude on its initial value $x_{0}$ disappears in the long-time limit, as in the case of purely quadratic drag.\cite{Zeldovich} Comparison of the asymptotics to a numerical integration of the equation of motion is shown in Fig.~\ref{fig:plot3}. A \textsc{Python} implementation of the numerical approach is available in supplemental material.\cite{code} Due to the exponential decay at long times, it should be emphasized that the argument that terms with $\alpha>2$ are non-physical\cite{Quiroga} due to oscillations which decay too slowly only holds when a {\it single} damping term with $\alpha > 2$ is present. Additionally, it is known that distinguishing between (for example) linear and quadratic functions as the dominant drag term from simulated experimental data is not always possible.\cite{Hauko2} Experimentally, one might expect the difficulty in determining the drag force to be compounded by the potential existence of more than one power-law terms, perhaps with one much ``weaker'' than the other. It might prove worthwhile to explore the feasibility of using Eq.~(\ref{eq:atwodrag}) to fit the relative amplitudes rather than distinguishing between two pure, power laws.

The methods described in this note allow one to investigate the amplitude decay in oscillating systems with any type of power-law damping. This approach provides a concise theoretical complement to the experimental investigation of nonlinear damping in oscillating systems. In particular, one could create a fairly rich lab experiment with strong theoretical and computational components by using the TeachSpin Torsional Oscillator\cite{TeachSpin} which offers ways of producing linear ($\alpha = 1$), quadratic ($\alpha = 2$) and sliding ($\alpha = 0$) drag in a mechanical system with a nearly linear restoring force. Additionally, the TeachSpin apparatus allows one to combine these types of drag, and the results of this note could form the basis of a theoretical model for such a setup which has no exact solutions. Lastly, the availability of affordable and accurate accelerometers allows one to collect detailed data in oscillating systems which can be analyzed to reconstruct the effective drag force.\cite{Hinrichsen} The results outlined in this note could provide a simplified theoretical component for interpreting such experimental data.
 \begin{acknowledgments}
The author thanks Robert Hauko and the anonymous reviewers for constructive criticisms and valuable insights. The author also thanks to Joseph M. Starobin for illuminating discussions and for introducing him to Ref.~\onlinecite{Zeldovich}.\\[5ex]
\end{acknowledgments}


\begin{thebibliography}{11}
\bibitem{note3} The general case of this integral can be evaluated formally in terms of the hypergeometric function $_{2}F_{1}(a,b;c;x)$ as
\begin{eqnarray}
\int \frac{dx}{c x^{\alpha} + c' x^{\alpha'}} & = & \frac{x^{1-\alpha'}\, _{2}F_{1}\left(1,\frac{\alpha'-1}{\alpha'-\alpha};\frac{1+\alpha-2\alpha'}{\alpha-\alpha'}; -\frac{c}{c'}x^{\alpha-\alpha'}\right)}{c'\left(1-\alpha'\right)},\nonumber
\end{eqnarray}
but one can argue that this expression is not much in the way of progress toward a physically transparent expression.
\bibitem{note5} Since $b\dot{x}$ and $b'\dot{x}^{2}$ both have the same dimensions, $b$ and $b'$ differ by a factor of velocity. To compare the quantities one must introduce an appropriate factor of velocity as $b' \rightarrow \omega_{0}x_{0}b'$.
\end{thebibliography}
\end{document}